# Rebuildable biochronometer: inferences and hypothesis on eukaryotic timing system

Ming-Jia Fu

## Abstract

The biochronometers used to keep time in eukaryotes include short-period biochronometer (SPB) and long-period biochronometer (LPB). Because the circadian clock reflects the biological time rhythm of a day, it is considered as SPB. Telomere shortening, which reflects the decreasing of telomere DNA length of chromosomes with the increase of cell division times, can be used to time the lifespan of organisms, so it is regarded as LPB. It is confirmed that SPB and LPB exist in most eukaryotes, and it is speculated that SPB and LPB are closely related. In this paper, based on existing studies, it is speculated that SPB and LPB of most eukaryotes can be co-attenuated with cell division in the process of aging. Due to the attenuated phenomenon of key components in the biochronometers during the growth and development of organisms, the biochronometers attenuate with the aging. Based on existing research results, it is preliminarily determined that the biochronometers can be rebuilt in the co-attenuated process. When the key components of biochronometers are reversed and increased in the organism, it can lead to the reversal of biochronometers, which further leads to the phenomenon of biological rejuvenation and makes the organism younger. In addition, the rebuilding of biochronometers can also lead to the acceleration of biochronometers and the shortening of the original timing time of biochronometers, thus shortening the life span of organisms. The rebuilding of biochronometers includes the reversal of biochronometers, the truncation of biochronometers timing and Uncoordinated co-attenuation of biochronometer and so on. The reversal of the biochronometers, which leads to rejuvenation, can give us a whole new understanding of life expectancy to be different from anti-aging.

**Keywords:** Biochronometer, Circadian clock, Telomere length (TL), Riochronometer co-attenuation, Reversible biochronometer, Rebuildable biochronometer

## Introduction

In the long-term evolution of eukaryotes, the internal molecular machines that can measure time, biochronometers, have been formed. When the molecular machines works normally, it is able to provide the time rhythm and timing of eukaryotes in a short period (such as one day), as well as the timing and regulation of eukaryotes in a longer life cycle.

As for the research of biological timing system, it is more about the research of biological rhythm clock, that is, the research of circadian clocks. The 2017 Nobel Prize in Physiology and Medicine was awarded to three American scientists Jeffrey C. Hall, Michael Rosbash and Michael W. Young for their discoveries of the gene and molecular mechanism of circadian rhythm. The research on circadian clock is abundant in animals, plants and fungi, and circadian clock is more likely to reflects the rhythmic changes of eukaryotic individuals in a day [1]. There are also studies on the biomolecular system of long period for eukaryotes, that is, the study of chromosome telomeres, especially the mechanism of telomere formation and telomere shortening, which has been identified as the landmark study of eukaryotic life span [2]. For this, American scientists Elizabeth Blackburn, Carol Greider and Jack Szostak won the Nobel Prize in Physiology and Medicine in 2009. Therefore, I define these two molecular systems, that describe the time properties of eukaryotes, as the short-period biochronometer (SPB) and the long-period biochronometer (LPB), respectively.

The biochronometer of eukaryotes, whether SPB or LPB, will change with the process of life, and the biggest change may be the attenuation of biochronometer, that is, when organisms lose the original characteristics of biochronometer in the course of life, including the disorder of timing in SPB and the shortening of telomeric DNA in LPB, cells will tend to be senescent. Here, I not only described the relationship between SPB and LPB, but also speculated and explained the possibility of biochronometer reversal and rebuilding in the process of life, which will lead to a reversal of life or an acceleration of death.

## Eukaryotic timing system

In fact, there have been many excellent studies on eukaryotic timing systems, which are irreplaceable and have incomparable advantages in explaining the molecular mechanism of certain timing periods in eukaryotes. The timing system for eukaryotes is naturally present in organisms, which can time a day of organisms and the longer life cycle of organisms. Therefore, it can be summarized as two main types of eukaryotic timing systems, namely, SPB (one-day eukaryotic timing systems for individual organisms) and LPB (eukaryotic timing systems for the longer life cycle of individual organisms).

### Short-period biochronometer

The eukaryotic SPB is a circadian clock or a 24-hour circadian rhythm, that is reflected in the circadian oscillators of cells based on transcription/translation negative feedback loop [1, 3–8]. Therefore, it can be considered as a short-period circadian timing system. At present, it can be determined that the short-period timing system exists in almost most organisms, and the relevant studies have been conducted in animals, plants, fungi and eubacteria [1]. The basic working mechanism of circadian clock is the formation of circadian oscillator. In these studies, key components of the circadian clock negative feedback loop of transcription and translation have been identified. A number of these components become positive regulators, and other components are negative regulators [1, 9]. The circadian oscillator is composed of positive and negative regulators. The oscillator is the core of the circadian clock work, which can receive external signals, cause corresponding gene expression, and then maintain the circadian activity of the organism by controlling the output pathway of the clock signal [1, 4, 8].

The circadian clock of *Drosophila* has been studied earlier. The main components of the circadian oscillator are CRY (CRYPTOCHROME), CLK (CLOCK), CYC (CYCLE), TIM (TIMELESS) and PER (PERIOD). Two heterodimer proteins that regulate gene transcription play a key role in the basic working mechanism of circadian oscillator: One is CLK-CYC, a dimer of transcription factors CLK and CYC that act directly on DNA to promote transcription. Another is PER-TIM, a dimer of PER and TIM, which inhibits the function of CLK-CYC. The function of CLK-CYC is to promote the expression of a series of genes related to circadian clock behavior, including *per* and *tim*. All of these genes have a DNA sequence called E-box element at the promoter site, and CLK-CYC acts on the E-box sequence to initiate the expression of these genes. The expressed PER and TIM proteins first accumulate in the cytoplasm gradually, and at night, when the two proteins accumulate to a certain amount, they are transported to the nucleus to inhibit the activity of CLK-CYC, thereby inhibiting the expression of themselves and all CLK-CYC relevant downstream genes and reducing the amount of expression. The PER protein in the cytoplasm is gradually hydrolyzed, thus forming a transcription/translation negative feedback loop circadian oscillator that negatively feedback regulates gene transcription and translation with a cycle of 24 h. Thus, the core of the circadian clock is determined to be a circadian oscillator of gene expression [3, 10–16].

Similar mechanisms of circadian clock work exist in mammals [6, 17–20], plants [21–23] and fungi [1, 24–27], all of which have the transcription/translation negative feedback loop circadian oscillators. It should be noted that in mammals, the circadian clock includes the central clock existing in the suprachiasmatic nucleus (SCN) and the peripheral clocks in the peripheral tissues, and central clocks and peripheral clocks have the same circadian genes and the same molecular clockwork mechanism [28–31]. Core positive and negative regulators in the core oscillator of representative organisms are shown in Table 1. According according to existing studies, most SPBs have similar working mechanisms. SPBs of other organisms are not described here. The model of the simple core oscillator work is shown in Fig. 1.

**Table 1.** Core components in the core circadian oscillator of some organisms

| Organisms | Positive regulators | Negative regulators | Citation |
|---|---|---|---|
| *Drosophila melanogaster* | CLK (CLOCK); CYC (CYCLE) | PER (PERIOD), TIM (TIMELESS), CRY (CRYPTOCHROME) | [10–16, 32, 33] |
| mammals | CLOCK (Circadian locomotor output cycles kaput); BMAL1 (Brain and muscle arnt-like protein 1) | PER; CRY | [20, 34–37] |
| *Arabidopsis thaliana* | TOC1 (TIMING OF CAB EXPRESSION 1)♯; LWDs (LIGHT-REGULATED WDs); TCP20; TCP22; RVEs; LNKs | LHY (LATE ELONGATED HYPOCOTYL); CCA1 (CIRCADIAN CLOCK ASSOCIADED 1) | [21–23, 38–52] |
| *Neurospora crassa* | WC-1 (White Colar-1); WC-2 (White Collar-2) | FRQ (FREQUENCY); FRH (FRQ-interacting RNA helicase) | [1, 24–27, 53–60] |

♯ With the further study on the function of TOC1, it has been found that TOC1 is a transcription factor with transcriptional inhibition function, and TOC1 can directly inhibit the expression of CCA1 and LHY genes [44, 61, 62].

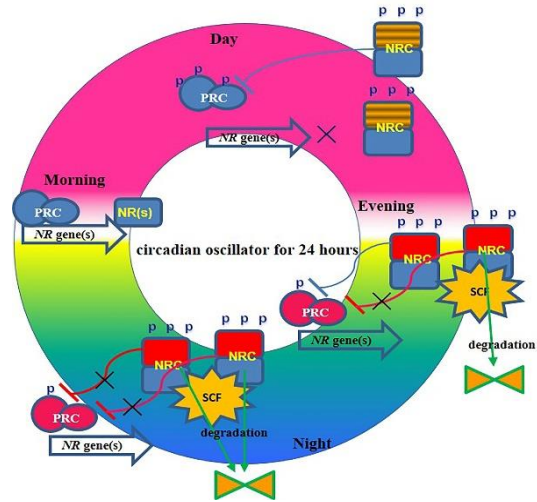

**Fig. 1** Proposed model of the short-period biochronometer (core circadian oscillator). In this model, the working principle of the core circadian oscillator is briefly described. In the morning, PRC (or PR) binds to *NR* gene(s) to promote the expression of *NR*(s); With the increase of NR(s) synthesis, NRC is formed in the daytime and the function of PRC (or PR) is inhibited. In the evening, NRC begins to ubiquitination and degradation, gradually removing the inhibition on PRC (or PR), and NRC degradation reaches a peak in the late night, and gradually starts to combine with NR gene(s) to start a new cycle. PR, Positive regulator. PRC, Positive regulator complex. NR, Negative regulator. NRC, Negative regulator complex. SCF, Skp1-cullin-F-box protein. P, Phosphate group.

## Long-period biochronometer

LPB can be considered as the deoxyribonucleic acid (DNA) synthesis-degradation biological clock, mainly related to the synthesis, degradation and modification of chromosomal DNA (more specifically, telomeric DNA), and is a biological clock related to the length of life of organisms. The longevity of organisms inevitably involves the long-term timing of individual organisms. There have been a lot of related researches and opinions, and some researchers have put the nuance of telomeres as a biological clock，and put forward the concept of mitotic clock [63, 64]. It has also been suggested that telomere length (TL) has been recognized as one of the best biomarkers of aging for a long time [2]. When TL is shortened to the point where DNA is damaged, cells are unable to compensate for this reduction in length, and chromosomal stability deteriorates, ultimately leading to cell death and body aging [65, 66].

Telomeres are the DNA structures enclosed by various proteins at the ends of chromosomes. The telomere sequence of mammals including humans is TTAGGG/AATCCC [67–69]. The telomere DNA sequences of different eukaryotes are very conserved in evolution, rich in G/C, and composed of a single type of base sequence. The number of repeats and the length of telomere sequence vary from species to species [70–76].

The stability of telomere is mainly related to the structural characteristics of telomere. In humans, the telomere DNA sequence is bound to shelterin, a specific complex consisted of six proteins. The protein complex consists of TRF1 (telomeric repeat binding factor 1) , TRF2 (telomeric repeat binding factor 2),  TIN2 (TRF1-interacting nuclear protein 2), POT1 (protection of telomeres 1), TPP1 (the POT1 and TIN2 organizing protein) and RAP1 (repressor activator protein 1) [77]. TRF1 and TRF2 in the shelterin complex directly recognize TTAGGG telomeric repeats in single-stranded and double-stranded DNA.  POT1 and TPP1 act as heterodimers and interact with the prominent single strand at the 3'-end of the telomere. TIN2 can link POT1/TPP1 heterodimer to TRF1 and TRF2, which can stabilize the interaction of TRF1 and TRF2 with telomeres [78]. RAP1 interacts with TRF2, which can increase the specificity of TRF2 binding to telomeric DNA and regulate the localization of TRF2 at the chromosome end [79, 80]. Shelterin can positively and negatively regulate the increase of telomere repeats through telomerase, and at the same time protect chromosome ends from being recognized as DNA damage by forming a lock-like chromosomal structure [81, 82]. The protruding single strand at the 3'-end of telomere folds into the double helix of telomere to form a "T-loop" structure, which forms a cap structure that prevents DNA damage at the end of chromosome. Shelterin complex maintains the stability of "T-loop" structure and also regulates telomerase activity at the end of chromosomes. The special structure of telomeres determines their special function: Telomeres protect the ends of eukaryotic linear chromosomes from DNA damage response, degradation, and heterochromatin structure due to abnormal recombination [72, 81, 83]. Under normal conditions, telomeres form T-Loop and G-quadruples complex (G4) secondary structures, which interact with shelterin complex to maintain the stability of telomere structure [84].

## Attenuation of biochronometer

Attenuation of biochronometer mainly refers to the weakening of timing ability of biological clock in the biological cells, which will lead to the accelerated aging or the physiological dysfunction.

### Attenuation of SPB during aging

The SPB attenuation can be defined as the gradual functional decline of its rhythmic biosynthesis, biological signal transmission and SPB itself in the process of biological growth and development, which is mainly manifested in the gradual disorder and imbalance of main components synthesis of SPB in the process of biological growth and development (Fig. 2). The input of external environmental signals and the output of rhythmic signals to maintain SPB are gradually weakened, until the survival of organism cannot be maintained and the organism is approaching the critical point of death (Fig. 2). In short, a robust circadian oscillator fades into a weak one.

At present, many researchers have studied the degeneration of fungi as the aging problem [85]. In the process of studying the circadian clock of *Cordyceps militaris*, I and my research partners found that the degeneration of *C. militaris* was closely related to the synthesis content of main components of circadian clock, CmFRQ (*C. militaris* FREQUENCY). Therefore, in our study, it was determined that the total content of CmFRQ synthesis in *C. militaris* during the aging process gradually decreased with time, which was reflected in the successive transfer culture process of *C. militaris*. The relative content of CmFRQ decreased with the increase of transfer times. The morphology of the fruiting body gradually deteriorated with subculture, that is, the characteristics of fruiting body deformity and short stature was appeared [86]. it is speculated that during the aging process of *C. militaris*, not only the main component of the circadian clock, CmFRQ, but also the function of the circadian clock itself is attenuated, which leads to the aging of *C. militaris* and the degeneration of fruiting bodies. Although the results were obtained in one fungus, they at least hint at other organisms.

In humans and rodents, the circadian clock can change with age [87]. As organisms age, many physiological processes lose their integrity, leading to a loss of resilience in the face of environmental challenges [88]. Aging reduces the amplitude and changes the phase shifting of the body's circadian rhythms, including sleep, body temperature, cortisol, melatonin and metabolism [88–93]. In addition, some studies have shown that the expression of core components of the circadian clock decreases with age. In peripheral blood cells, it was found that the expression level of BMAL1 was negatively correlated with female age [94]. In elderly mice, the rhythmic amplitude of PER2 decreases [95]. In the suprachiasmatic nucleus (SCN) of hamster and mouse, the expression of Clock and Bmal93 decreases with age [96, 97]. Mutations in the circadian clock gene can accelerate aging in fruit flies and mice [98–100]. There is still some evidence that the circadian clock is unbalanced or attenuated by the reduction or loss of the expression of clock-related genes as individuals age [101–105]. These results can also be understood as the physiological decline and rhythm attenuation of the circadian clock in the aging process.

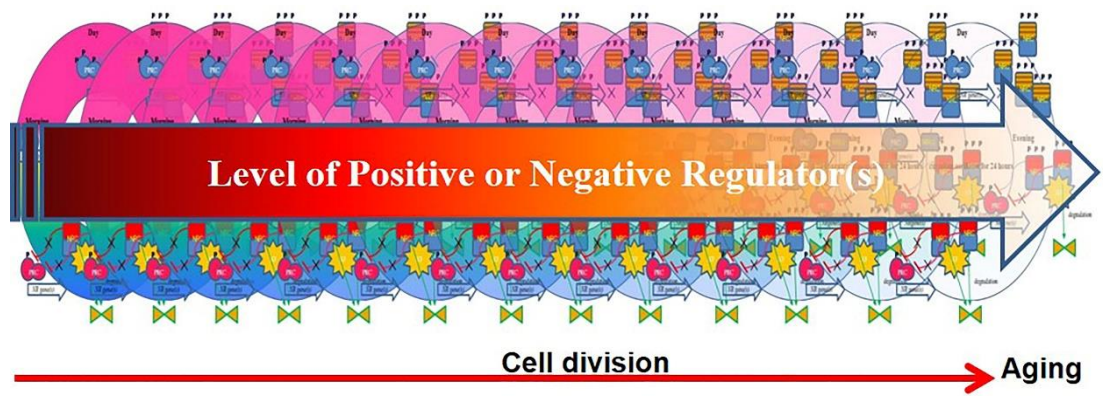

**Fig. 2** Proposed model for the attenuation of SPB. In organisms, with the increase of cell division times, cells gradually tend to age. In the process of biological growth and development, SPB in cells gradually attenuates with the increase of cell division times (Each circle in the figure represents a circadian clock negative feedback loop of core circadian oscillator, and the color of each circle gradually faded from left to right represents the attenuation of these feedback loops function). Moreover, the level of major positive or negative regulators in the core circadian oscillator may gradually decrease with cell division (The arrow in the middle of the figure represents the level of positive or negative regulators in the cell, and the color of arrow gradually faded from left to right represents the level of positive or negative regulators gradual decreases).

## Attenuation of LPB during aging

LPB has been determined to time the life span of living things brought by the degradation of telomeric DNA, so the attenuation mechanism of LPB is also the degradation mechanism of telomeric DNA. At present, it is known that human telomere DNA will gradually shorten during cell division, and incomplete replication of chromosome ends will lead to telomere shortening, eventually leading to telomere dysfunction and irreversible cell cycle arrest, also known as replicative senescence [106]. Every time human cells divide, telomeres shorten by 50-100 bp [65]. Replicative senescence is induced by telomere shortening, which eventually reach the critical length at which the cells can either undergo apoptosis or stop dividing, thus entering a state of senescence [2, 107]. Telomere shortening limits somatic cells to a finite number of cell divisions, and the cell cycle is blocked between G1 and G2/M phases and gradually moves towards death [108–115]. Therefore, telomere shortening can be considered as a attenuation process of LPB (Fig. 3).

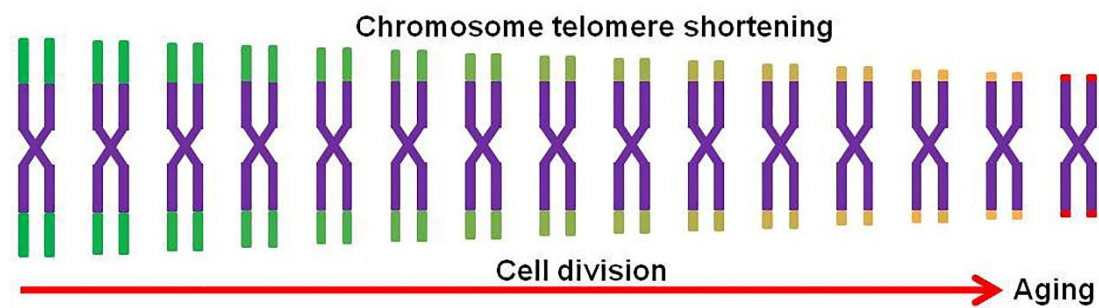

**Figure 3.** Proposed model for the attenuation of long-period biochronometer. The telomere theory holds that telomeres shorten with mitosis. When telomeres shorten to a certain critical length, cells lose the ability to proliferate. In the figure, telomeres on chromosomes are gradually shortened with cell division. The color of telomeres changes from green to orange and finally to red, indicating the shortening of chromosomes, that is, the attenuation process of LPB.

## Speculation of co-attenuation mechanism of SPB and LPB

In the aging process of eucaryotes through life, there are always SPB and LPB in the process of attenuation. Therefore, it is speculated that there is a connection between SPB attenuation mechanism and LPB attenuation mechanism in cells. At present, it is impossible to determine the signal pathways associated with the two attenuation mechanisms, but it can be speculated that there should be the correlation factors and signal pathway in the process of cell senescence, which can jointly regulate the attenuation process of SPB and LPB.

Studies have shown that SPB (circadian clock) can periodically output signals. The key regulatory factors of circadian clock output the rhythmic signals of circadian clock by regulating its downstream *clock-controlled genes* (*ccgs*), so that molecular activities in cells also show a time rhythm [116]. In mammals, the positive regulator complex CLOCK:BMAL1 binding occurs mainly at consensus E-box DNA motifs (DNA sequence CACGTG), which exists in the negative regulator genes *Per* and *Cry*, and as well as *ccgs* [34, 37, 117, 118]. Analyzing the enhancer of mammalian telomerase reverse transcriptase (TERT) gene, E-box sequences can also be found [119]. Therefore, TERT gene can be regarded as *ccg*. The circadian expression of circadian clock genes and *ccgs* maintains a high degree of synchronization and order of normal physiological activities [120, 121]. *ccgs* include the key regulator genes of tumorigenesis, cell cycle, cell proliferation, metabolism, senescence and DNA damage response [122–130]. Study also show that the circadian regulators are related to the key factors controlling chromatin remolding [131, 132], and the many key factors in DNA replication, recombination and repair have been identified as *ccgs* in mice [131, 133–136]. The circadian cycle and cell cycle can be coordinately control. The circadian clock regulates the key phases of the cell cycle, while the cell cycle is a strict regulatory system with checkpoints, precisely regulating mitosis and DNA replication [137–140]. The circadian clock participates in telomere homeostasis by regulating telomerase activity, TERT mRNA level, telomeric repeat-containing RNA (TERRA) expression and the circadian rhythm of telomere heterochromatin formation, and disruption of normal circadian rhythms accelerates aging and corresponds to shortening of telomeres [141–143]. Therefore, it is speculated that the circadian clock directly or indirectly regulates chromosome and DNA replication during cell division by regulating *ccgs*, thus regulating telomere DNA replication. It can be assumed that some factors in SPB will attenuate with cell division (more importantly, their expression will be disordered), while key factors in LPB will also attenuate, and SPB attenuation factors and LPB attenuation factors will interact. The resonance phenomenon of attenuation for SPB and LPB is occured, resulting in the simultaneous attenuation of SPB and LPB (Fig. 4).

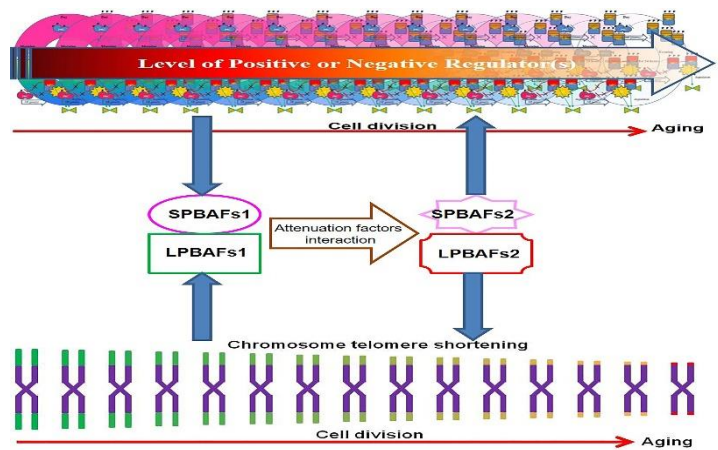

**Fig. 4** Proposed model for the co-attenuation of SPB and LPB. It is speculated that both SPB and LPB will attenuate in cells, and the regulatory factors of SPB may directly or indirectly induce the initial attenuation factor SPBAF1; LPB may also directly or indirectly induce the attenuation factor LPBAF1. The interaction between SPBAF1 and LPBAF2 leads to the generation of new signal factors SPBAF2 and LPBAF2, and the processed attenuation signals are fed back to SPB and LPB respectively, so that SPB and LPB further resonate and attenuate harmonically. SPBAF: short-period biochronometer attenuation factors; LPBAF: long-period biochronometer attenuation factors.

## Reversing and rebooting the biochronometers give the rejuvenescence to organisms

In most people's cognition, the physical time is not reversible, that is, the present time and the future time can not be returned to the past, at least the current science can not do time shuttle back to the past. It seems that the biochronometer used to keep time are also irreversible, and organisms can only age and die over time, putting a huge limit on how long they can live. The eukaryotic timing system obviously also goes awry during the aging of organisms. Specifically, the expression of the main components of SPB gradually disordered, leading to the attenuation of SPB. The main component of LPB is chromosomal telomeric DNA, which gradually gets shorter as cells divide. However, when examining the timing systems of eukaryotic organisms, it can be determined that the timing systems of eukaryotic organisms are composed of a variety of biomacromolecules, and the attenuation process of biochronometers is also the process of disordering the biosynthesis of biomacromolecules. Conversely, could this attenuation be prevented or even reversed if the biomacromolecules that make up the biochronometers in the cell were correctly synthesized in large quantities again? The answer is yes, the biochronometer can be reversed if the synthesis of its main components in an organism can be reversed by means of some means. This can also causes the aging process of the organism to be reversed and the organisms to be returned to a certain stage of youth.

It can be preliminarily known that the attenuation of biochronometers is the process of biological aging. So it's conceivable that slowing down the biochronometers could extend life span. It's tempting to think of a more daring idea: Turning the biochronometers back to reverse the aging of life. If so, the aging of life can be prevented, or made younger. Such a scenario is certainly exciting and desirable, and can lead to new areas of research into biological life span. Now we have a better understanding of the mechanism of the biochronometers, whether it is SPB or LPB, they are composed of biomacromolecules. On the basis of biomacromolecules composition, if there is a mechanism to promote the massive synthesis of the constituent biomacromolecules of the biochronometers, it will surely reverse the attenuation of biochronometers, in effect, the biochronometers will be reversed to make organisms younger.

In the aforementioned study on the degeneration of *C. militaris*, when the degenerated *C. militaris* were treated with hypertonic solution and proteasome inhibitor MG132, the massive synthesis of CmFRQ in the mycelium was increased, and the fruiting body was rejuvenated [86]. Therefore, it was analyzed that when *C. militaris* was treated with hypertonic solution, mycelium growth was slow, but CmFRQ, a key factor of circadian clock, was accumulated. Once the influence of hypertonic solution is removed, the content of CmFRQ remains at a high level, leading to the reversal of SPB, so *C. militaris* appeared to be rejuvenated. Moreover, MG132 treatment can inhibit the degradation of proteins including CmFRQ in cells, and can also lead to the reversal of SPB and the rejuvenescence of *C. militaris*. Therefore, a model of biochronometers reversal is proposed here (Fig. 5). In this model, SPB and LPB will attenuate harmoniously with the growth and development of the organisms. However, the treatment of slowing down cell division and inhibiting protein degradation in the early stage, early middle stage and later middle stages of biological growth and development can lead to the accumulation of positive or negative regulators of short-period biochronometer, thus reversing the short-period biochronometer (Fig. 5). It is also conceivable that using the same method would lead to a reversal of LPB. The reversal of the biochronometers eventually causes an organism to be rejuvenated. But in the senescent cells, biochronometer cannot be reversed.

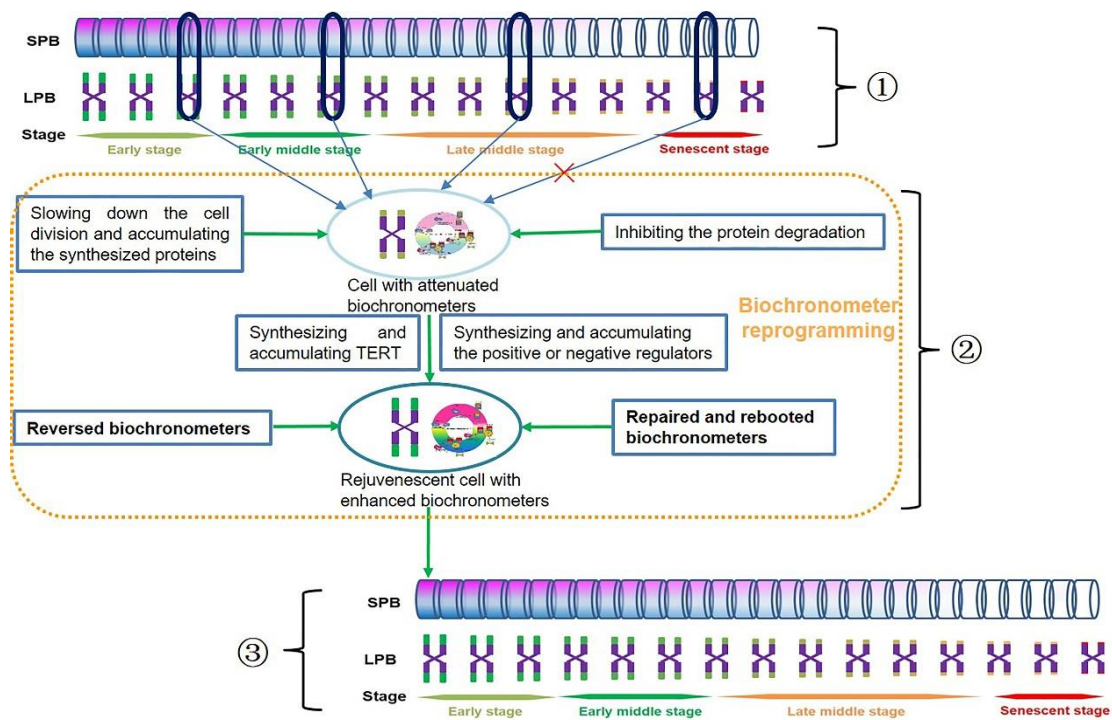

**Fig. 5** Proposed model of biochronometer reversal. ① In the process of biological growth and development, SPB and LPB co-attenuate at different growth and development stages. In the early stage, the biochromomers does not co-attenuate significantly. In the early middle stage and late middle stage, the biochromomers gradually co-attenuates, and in the aging stage, the biochromomers significantly co-attenuates; ② When the cells from early stage, early middle stage, and later middle stage were treated to slow down cell division and inhibit protein degradation respectively, TERT, PR, and NR were accumulated in cells, and SPB and LPB were reversed. During this process, the biochronometers are reprogrammed. However, SPB and LPB in the senescent cells could not be reversed; ③ After the reversal of SPB and LPB, the biological cells are rejuvenated, and the individual organisms become younger and re-enter the new growth cycle of organisms. SPB, short-period biochronometer; LPB, long-period biochronometer;

TERT, telomerase reverse transcriptase. PR, Positive regulator; NR, Negative regulator.

## Rebuilding of biochronometer

In the biochronometers of normal somatic cell, SPB and LPB can be co-attenuated, and the mechanism of such co-attenuation is mainly caused by the attenuation or reduction of key factors in SPB and LPB. Therefore, if the content of key factors in SPB and LPB is changed in cells, the corresponding timing phase of the cells in which SPB and LPB are located will be inevitably changed, that is, the growth and development stage of the cells will be changed, leading to the rebuilding of biochronometers. The aforementioned reversal of biochronometers is a special example of biochronometer rebuilding. The rebuilding of biochronometers can lead not only to younger living organisms, but also further aging or disease. Three particular examples are discussed below to illustrate the unique timing characteristics of living things brought about by the rebuilding of biochronometers.

## Co-attenuated biochronometers of truncated timing phase in cloned sheep Dolly's cells

Dolly, a cloned sheep, was born on July 5, 1996 [144]. The research result were shocking at that time, and seemed promising for asexual reproduction of mammals. However, Dolly had health problems in her life, and her short life ended in just six years [145, 146]. Dolly's early death brought more thinking to people [147]. Dolly's cell nucleus cames from a 6-year-old Finland Dorset sheep [144]. The somatic cell of an adult sheep has divided many times, but Dolly's life started from this senescent nucleus, despite the fusion of another ovum cell without a nucleus. It was found that the telomere length in Dolly's cells was shorter than that of normal sheep of the same age, and showed the signs of premature aging such as senile arthritis [146].

By analyzing the origin of the biochronometers in Dolly's initial cells, it can be determined that LPB came from the nucleus of the somatic cell that has divided many times, while SPB came from the cytoplasm of the ovum that has not divided yet. This kind of biochronometers after cell fusion has an unbalanced match between LPB and SPB. LPB has experienced a deeper attenuation, while SPB has not yet shown attenuation. If there is a co-attenuation mechanism between LPB and SPB, SPB will also attenuate synchronously with LPB in the new fusion cell, and the individual growing from this cell will enter the stage of LPB. This is exactly what happened to Dolly's growth, as she entered middle age from the start, and brought SPB into sync with LPB (Fig. 6). As a result, Dolly's biochronometers had been reprogramed and rebuilt by the time she was born.

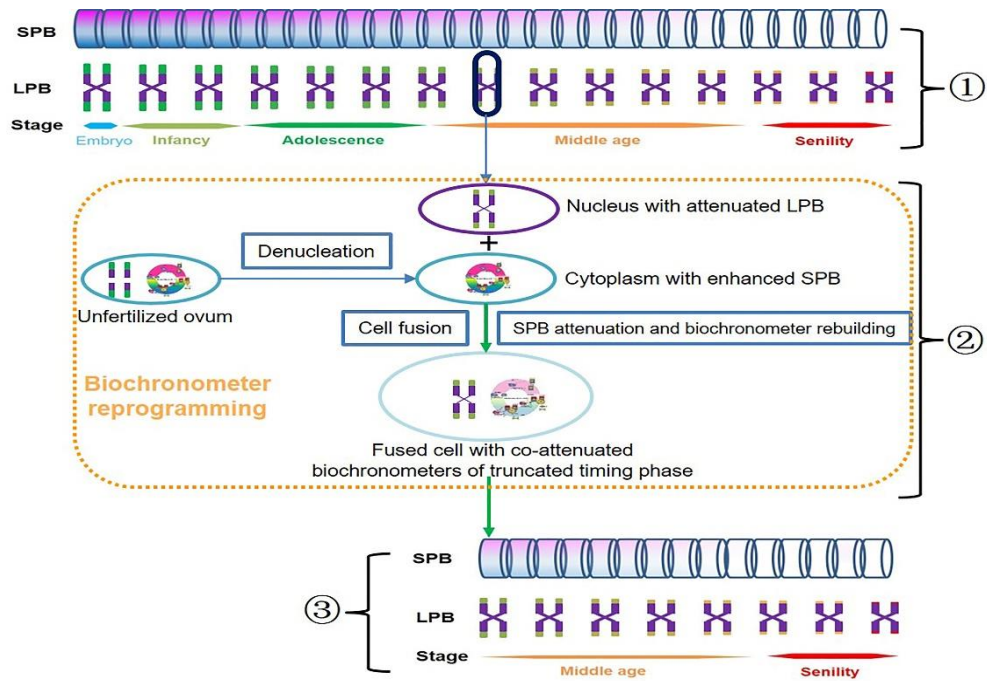

**Fig. 6** Co-attenuation of biochronometer of truncated timing phase in cloned sheep Dolly's cells made her directly enter the middle age stage. ① In the process of growth and development, SPB and the LPB are co-attenuated at different growth and development stages; ② When the nucleus of the middle-aged somatic cell fuses with the enucleated unfertilized ovum, the nucleus of the middle-aged somatic cell has attenuated LPB, while the cytoplasm of unfertilized ovum has enhanced SPB. After cell fusion, there is the co-attenuated biochronometers in the fused cells; ③ After the cells with co-attenuated LPB and SPB develop into biological individuals, they are in the development stage of the original LPB, and the subsequent life cycle is shortened.

## Uncoordinated attenuation of biochronometers in mammalian cancer cells

Present findings have shown that the disruption and dysfunction of circadian clock can contributes to the development of a variety of cancers [120, 148–150]. In cancer cells, the expression of core circadian regulators is in the loss of homeostasis. Whether the expression of the core circadian regulators decreases or increases, it can induce the risk of cancer cell occurrence. The tumor tissues in pancreatic ductal adenocarcinoma patients shows significantly lower PER-1, PER-2, PER-3, CRY-2, and CK1ε expression level [151]. The expression of PER-1, PER-2, and PER-3 is low in non-small cell lung cancer(NSCLC) and breast cancers [152, 153]. In hepatocellular carcinoma tissue, the expression levels of PER-1, PER-2, PER-3, CRY-2, and TIM were decreased [149]. The overexpression of PER-2 inhibits the migration and invasion of NSCLC cell lines [153, 154]. BMAL1 and PER2 were found in malignancy breast cancer cells, but maintained at a very low level [155]. The fact that the decreased expression and polymorphism of the core circadian genes *Per1, Per2* and *Per3* appear in a variety of cancer tissues have been reviewed by Fu and Kettner [120]. The circadian clocks in human cancer cells are characterized by deregulation or polymorphism of multiple or all core circadian genes [120]. Therefore, it can be determined that SPB in cancer cells is in disorder, and the expression of core components is not balanced, leading to the continuous attenuation of core oscillator of SPB.

Normally, TL declines with age in somatic cells, which result in an obstacle to tumor growth and contribute to cell loss with age [156, 157]. To counteract this shortening, certain types of cells express telomerase, which has the potential to restore shortened telomeres. These cells belong to highly proliferative cells, including germline cells, granulosa cells, early embryos, stem cells, activated lymphocytes, hematopoietic cells, basal epidermal cells, immortal cells, and certain types of cancer cell populations [158–161]. In tumor cells, there are a variety of telomere maintenance mechanisms, of which the action of telomerase is the most extensive mechanism of telomere extension [162]. Therefore, in tumor tissues, the main feature of cells is continuous division but no shortening of telomeres, which reflects the continuous non-attenuation characteristic of LPB.

Analysis of SPB and LPB in cancer cells shows that the two biochronometers are out of sync. When a normal cell transforms into a cancer cell, the biochronometers is rebuilt, with SPB shifting to a state of continuous attenuation, while LPB becomes a state of continuous non-attenuation due to the non-shortening of telomeres (Fig. 7). Although cancer cells have become immortal cells that can continue to divide, their functions are not comprehensive due to the disharmony of two biochronometers. So cancer cells can also be thought of as cells whose biochronometers are out of whack.

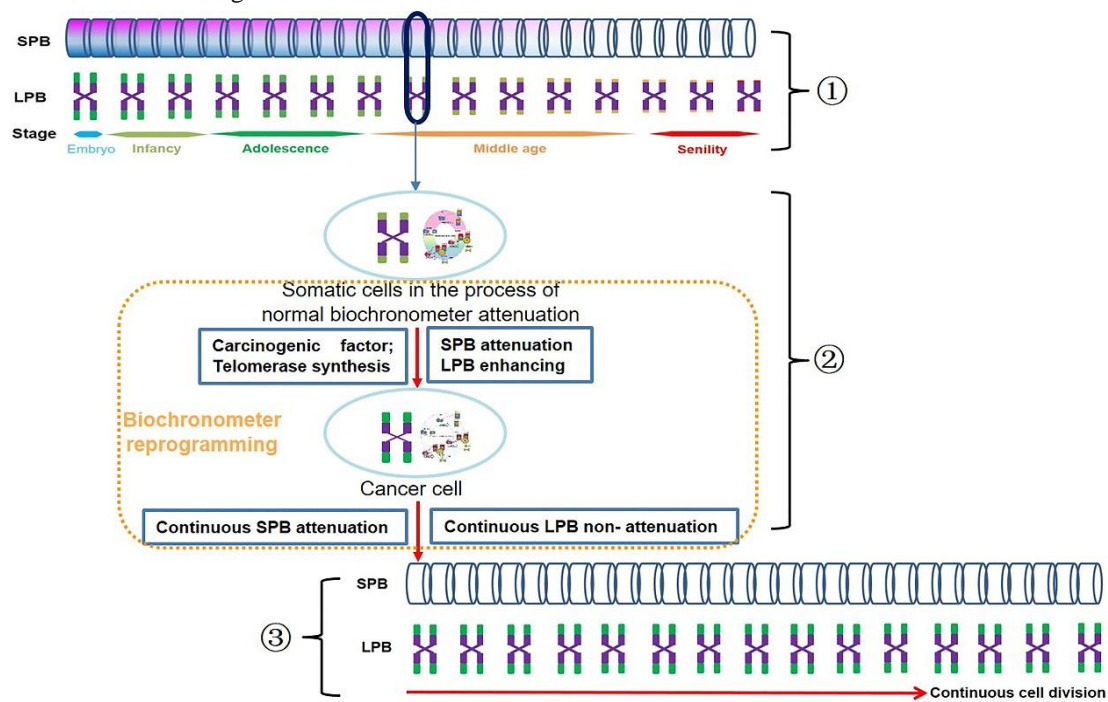

**Fig. 7** When normal cells were transformed into tumor cells, SPB and LPB changed from balanced co-attenuation state to unbalanced continuous attenuation of SPB and continuous non-attenuation of LPB. ① In the process of normal cellular growth and development, SPB and the LPB are co-attenuated at different growth and development stages; ② Under the influence of carcinogenic factors and the synthesis of telomerase, the normal attenuation mechanism of SPB is destroyed, and the telomere is not shortened and enhanced due to the continuous synthesis of telomerase; ③ In cancer cells, mismatched SPB and LPB are reflected in the maintenance of unshortened telomere DNA and the continuous attenuation of SPB, and the cancer cells can continue to divide.

## Rebuilding and rebooting of biochronometers in mammalian

The spermatogenesis and oogenesis of mammalian go through the stages such as spermatogonium and oogonium (2n), primary spermatocyte and primary oocyte (2n), secondary spermatocyte and secondary oocyte (n), sperm and ovum (n). After that, the development goes through the stages such as fertilized ovum, 2-cell, 4-cell, 8-cell, morula, blastocyst, gastrula, infant, adolescence, middle age and old age. When the offspring of organisms start a new life, it is logically impossible for them to continue their parents' biochronometers, and they must use an updated biochronometers in order to have a longer lifespan. Therefore, the rebooting of life undoubtedly requires the rebuilding and reversal of the biochronometers. Obviously, the rebuilding of this biochronometers occurs primarily at the beginning stage of individual organisms.

Here, it is necessary to analyze the changes of SPB and LPB in the production of mammalian germ cell genesis and the development of fertilized ovum into a mature individual, which is conducive to determining the main factors related to the reversal of the biochronometers in the rebuilding of the biochronometers.

The germ cell genesis of mammalian involves specific morphogenesis, changes and interactions of biomacromolecules, hormonal control, epigenetic reprogramming and other spermatogenesis and ovogenesis processes to form haploid sperm and ovum [163–171]. The expression of circadian clock genes in spermatogenesis and oogenesis is significantly different from that in somatic cells. The expressions of clock genes *bmal1*, *clock*, *cry1*, *per1*, and *per2* have been confirmed in human sperm [172]. Both BMAL1 and CLOCK contribute to the CB (chromatoid body) assembly and physiology during spermatogenesis [173]. The expression of circadian clock genes *Bmal1* and *Per1*, *Per2*, *Clock*, *Cry1*, and *Npas2* in the testis of mice remains constant within 24 hours, and the expression of PER1 and CLOCK proteins is restricted in cells at specific developmental stages of spermatogenesis, indicating a lack of functional circadian rhythms [174, 175]. The circadian genes *clock*, *bmal1*, *cry1*, *cry2*, *per1* and *per2* are expressed in the mouse oocytes and preimplantation embryos and function as maternal mRNA regulatory events. Circadian genes do not involve in the regulation of circadian clocks in mouse, cow, rabbit oocytes and preimplantation embryos, but in physiology, such as meiosis [176, 177]. As a result, the circadian clock (SPB) has a limited functional role in germ cell genesis and early embryo development, and does not rhythmically regulate biological processes or has no rhythmic output at all. Therefore, it is speculated that this process is a process of updating and strengthening the components of SPB.

The length of telomere repeats reflects the life span of organisms and represents LPB. Therefore, the complete telomere repair is undoubtedly necessary in the early stages of life, which is mainly reflected in the changes in telomerase levels at different stages of germ cell formation and embryonic development. The formation and maintenance of telomere repeats require the function of telomerase, so the emergence and possession of telomerase function in cells is a key event for telomere rebuilding. There are some differences in the studies on the expression of telomerase in different developmental stages of mammals among researchers, but it can generally reflect the change level of telomerase in different developmental stages of mammals. TL increased during the development of male germ cells from spermatogonia to spermatozoa and was negatively correlated with the activity of telomerase [178]. The telomerase activity is inhibited in the differentiated spermatozoa [179, 180]. High telomerase activity is detected in blastocysts, but not in mature sperm or oocytes during human embryonic development [181]. Human telomerase activity can be detected in all stages of early development from oocytes to blastocyst stage, reaching its highest level in morula and blastocyst stage and then decreasing in inner cell mass stage. In fetuses, telomerase is expressed only in tissue-specific stem cells at the end of embryonic period and organogenesis. Telomerase activity is down-regulated in postnatal somatic cells [182, 183]. Many researchers believed that telomerase activity exists in early embryos, but its level varies with different developmental stages. The telomerase activity of mature oocytes was relatively low. It increases after in vitro fertilization and then gradually decreases until the embryo reaches the eight-cell stage. Telomerase activity increased again after the eight-cell stage and reached its highest level at the blastocyst stage [184–186].

Although many physiological and biochemical processes are involved in the rebuilding of mammalian biochronometers, the existing data can be used to infer the pattern of mammalian biochronometers rebuilding process (Fig. 8). In the process from spermatogonium and oogonium to sperm and ovum, the expression of circadian clock genes loses its original biological rhythm function, and the activity of telomerase decreases with the elongation of telomeres. It is speculated that the production of sperm and ovum in this period may lead to the removal of the harmful redundant components or the components of blocking the biochronometer work (bugs). Therefore, the process of eliminating "bugs" can be thought of as the process of "formatting" the biochronometers. In this process, the biochronometers from the mother or father will be modified, in which the original factors that can attenuate the SPB and LPB will be eliminated. During the biochronometer "formatting" process, the biochronometers is also reprogrammed. During development from the fertilized ovum to the early embryo (mainly during the morula period), the biochronometers continues to be reprogrammed. After entering blastocyst, the cells begin to differentiate, the reprogramming ability of the biochronometers decrease, and it enters a process of co-attenuation.

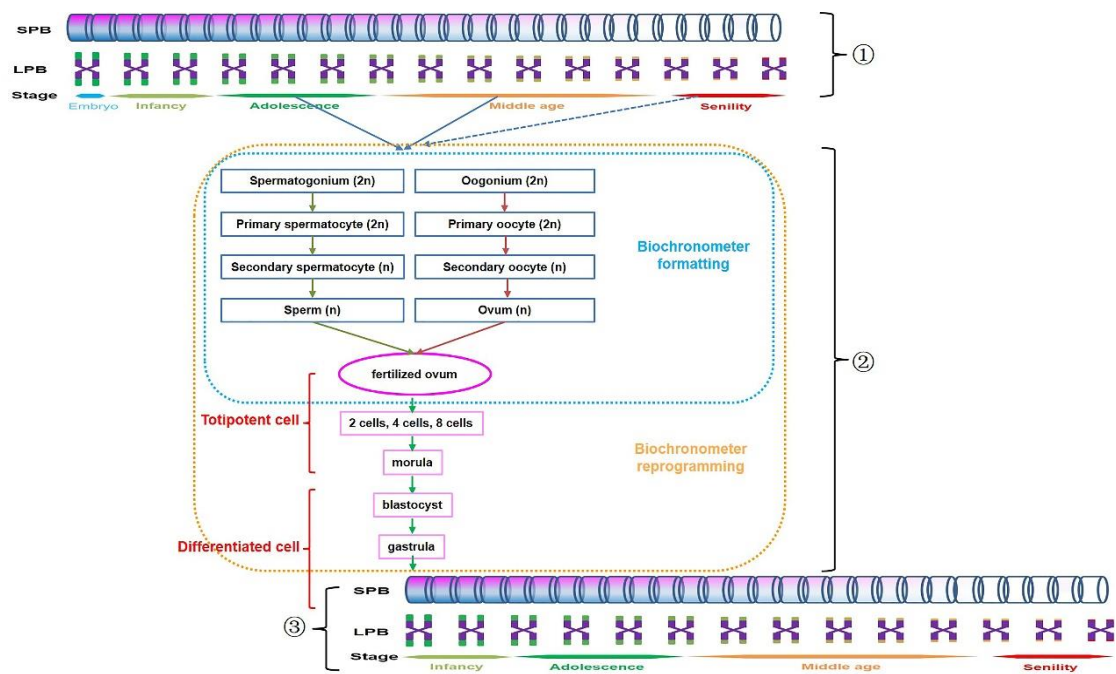

**Fig. 8** Proposed model of rebuilding and rebooting of biochronometers in mammalian. ① In mammalian parents, both SPB and LPB co-attenuate. ② Rebuilding of biochronometers in the offspring of mammals: In the process of spermatogenesis and oogenesis, according to the available information, it can be determined that the circadian regulator gene is expressed, but the circadian regulator has no rhythmic function of the circadian clock, only other biological functions; During the time period from spermatogonium and oogonium to sperm and ovum, telomere repeats are increased, but telomerase activity is decreased, and even no telomerase activity in sperm and oocytes. Therefore, it is speculated that the factors (bugs) that promote the attenuation of the biochronometers will be eliminated during the spermatogenesis and oogenesis, and the stage of clearing "bugs" can be considered as the "formatting" process of the biochronometers (represented by the blue dotted box). At present, it is uncertain whether the "formatting" process of the biochronometers will continue to the zygote, or even to the stage of the morula (which is still in the totipotent cell stage). Here, it can be hypothesized that epigenetic biochronometer reprogramming (indicated by the orange box) occurs during the period from the onset of spermatogonium and oogonium to the formation of sperm and ovum, and from fertilized ovum to the completion of embryonic development, resulting in the biochronometer being in its strongest non-attenuated state after fetal birth. ③ The offspring of mammals began to run the biochronometer, and new life entered the co-attenuation process of the biochronometer again.

## Disscussion

The circadian clock and telomere shortening in somatic cells are regarded as SPB and LPB, respectively, in an attempt to correlate the two types of biochronometers. It is inevitable that SPB and LPB is co-attenuated, and its significance lies in the fact that the periodic rhythmic behavior in cells also rhythmically regulates the length of life. The available data do link the two types of biochronometers, mainly by the fact that the key regulator of SPB can act directly or indirectly on cis-acting element of many protein factor genes involved in DNA replication and telomere synthesis or telomere shortening.

There are obvious differences in co-attenuation among different organisms, which exist in the work process of SPB and LPB of different species. At present, there is little data on the differences of SPB attenuation among different species, and the changes of TL in different species have been studied in detail. The correlation between TL and age has been well established in animals, and in many animals telomere length decreases with age [67, 112, 187–192]. TL can be used as molecular clocks or cellular timekeepers related to lifespan, reflecting the age level of an individual organism to some extent. Thus, in many animals, SPB and LPB have a typical co-attenuation mechanism. However, the variation of TL in plants was regulated by telomerase activity and varied with the increase of physiological age. The variation of TL was different under different plant species, different tissue parts and different environmental conditions [193–198]. Unlike many animals that only have telomerase activity at specific periods, studies have shown that a variety of plant telomerase can be distributed in meristem and organs with relatively vigorous cell division [191], and high telomerase activity was also found in callus [189]. Due to the diversity of TL in plants and the fact that TL does not necessarily shorten with age, there is a telomere maintenance mechanism in plants, which is different from the rule that TL gradually decreases with aging in many animals. For maintenance of TL, in addition to telomere repeat sequence extension through telomerase, there is another telomere replacement extension pathway with homologous recombination [199]. Loss of telomerase and canonical telomere repeats occurs in some plants and animals, especially in Diptera and many species of Coleoptera and Hemiptera [200]. This also suggests that the function mode of LPB may not be completely the change of TL, or there may be other modes. The understanding for existing organisms is that there is no immortal biological species, and all organisms have a limited lifespan. Therefore, both SPB and LPB will experience attenuation over a longer time range. It is speculated that after a long period of time, the mechanism for maintaining SPB and LPB will encounter errors that are difficult to be reversed. Eventually, SPB and LPB will co-attenuate, but this co-attenuation will become more complex. The co-attenuation mechanism of SPB and LPB can also be further regarded as the co-attenuation of the maintenance mechanism of SPB and LPB.

For the most part, the biochronometers continues to run in the same direction. However, when the working conditions of biochronometers are broken, the original operating trajectory will also change rapidly. For example, when physiological lesions occur in organisms, the key regulatory factors in both SPB and LPB can be changed, leading to the rebuilding between SPB and LPB, which may accelerate the process of aging. The most obvious example of reversing the biochronometers is in the fertilized ovums of many animals, where the biochronometers are rebuilded and reversed, leading to a rebooting of life (Fig. 8). If the rebuilding of biochronometers leads to unbalance or rapid attenuation of SPB and LPB, it will lead to biological disease or premature aging. Thus, the biochronometer rebuilding can be multi-dimensional or multi-directional (Fig. 9). Apparently, the different working results of biochronometers have led to the rebuilding of the biochronometer, and led to normal aging, premature aging, and disease in life. The better result is to rejuvenate the organisms into a more youthful state.

The life span of organisms is limited, and the process of life is also the process of aging. So the study of aging is an old and young topic. da Costa et al. reviewed a variety of aging theories, including program theories, damage theories, and combined theories [201]. They also describes the changes that occur during aging, including molecular changes, physiological changes, pathological changes, and psychological changes. Several options have also been proposed for current aging therapies. No matter what kind of method, it can't bring the biological individual back to the young state. Aging is the gradual deterioration of bodily functions over time. Taking an anti-aging approach only delays the aging process, but it does not make the organisms any younger. Reversing the biochronometers, however, allows the organisms to improve and return to a state of youth, or to rejuvenate. Therefore, the rejuvenation brought about by the reversal of the biochronometers is different from anti-aging, and will bring a new imagination space for life survival and longevity extension.

The rebuilding of biochronometers is a research area that is not fully understood at present, and will also bring us a new understanding of life. Two directions can be developed by the rebuilding of biochronometers: on the one hand, through the rebuilding of biochronometers, the aging or diseased cells can be rehabilitated; On the other hand, by rebuilding of biochronometers, healthy cells could also turn into unhealthy ones, which often happens in the human body, such as the canceration of normal cells. So far there has been no theoretical progress in the rebuilding of biochronometers and the transformation of cells in a better direction. Therefore, we need to strengthen our research work in this area in the future. Perhaps this will be a new area of research, but it will certainly give mankind more hope for health.

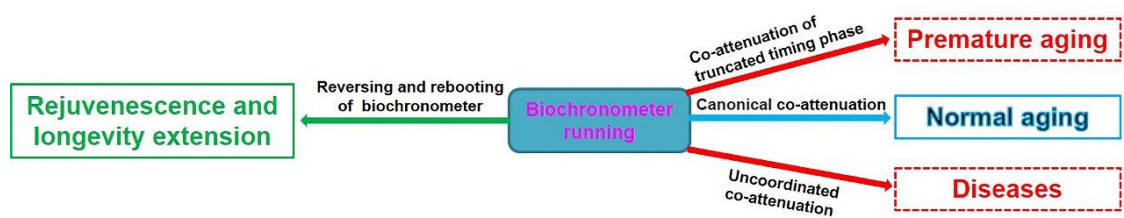

**Fig. 9** The rebuilding of biochronometers may bring different results to organisms.

**Funding**
The National Natural Science Foundation of China (Grant Number 31760601).

**Availability of data and materials**
Not available.

**Declarations**
**Ethics approval and consent to participate**
Not applicable.
**Consent for publication**
Not applicable.